\documentclass[conference]{IEEEtran}
\IEEEoverridecommandlockouts
\usepackage{cite}
\usepackage{amsmath,amssymb,amsfonts}
\usepackage{algorithmic}
\usepackage{graphicx}
\usepackage{textcomp}
\usepackage{xcolor}
\def\BibTeX{{\rm B\kern-.05em{\sc i\kern-.025em b}\kern-.08em
    T\kern-.1667em\lower.7ex\hbox{E}\kern-.125emX}}

\usepackage[T1]{fontenc}

\usepackage{url}
\usepackage{pgfplots}
\pgfplotsset{width=7cm,compat=1.3}
\usetikzlibrary{pgfplots.groupplots,pgfplots.dateplot,patterns}

\makeatletter
\pgfplotsset{
	groupplot xlabel/.initial={},
	every groupplot x label/.style={
		at={($({\pgfplots@group@name\space c1r\pgfplots@group@rows.west}|-{\pgfplots@group@name\space c1r\pgfplots@group@rows.outer south})!0.5!({\pgfplots@group@name\space c\pgfplots@group@columns r\pgfplots@group@rows.east}|-{\pgfplots@group@name\space c\pgfplots@group@columns r\pgfplots@group@rows.outer south})$)},
		anchor=north,
	},
	groupplot ylabel/.initial={},
	every groupplot y label/.style={
		rotate=90,
		at={($({\pgfplots@group@name\space c1r1.north}-|{\pgfplots@group@name\space c1r1.outer
				west})!0.5!({\pgfplots@group@name\space c1r\pgfplots@group@rows.south}-|{\pgfplots@group@name\space c1r\pgfplots@group@rows.outer west})$)},
		anchor=south
	},
	execute at end groupplot/.code={%
		\node [/pgfplots/every groupplot x label]
		{\pgfkeysvalueof{/pgfplots/groupplot xlabel}};  
		\node [/pgfplots/every groupplot y label] 
		{\pgfkeysvalueof{/pgfplots/groupplot ylabel}};  
	}
}

\def\endpgfplots@environment@groupplot{%
	\endpgfplots@environment@opt%
	\pgfkeys{/pgfplots/execute at end groupplot}%
	\endgroup%
}
\makeatother

\usepackage{booktabs}
\usepackage{arydshln}
\usepackage{changepage}

\usepackage{flushend}

\begin{document}

\title{Public Perception of the German COVID-19 Contact-Tracing App \emph{Corona-Warn-App}\\
\thanks{The COMPASS project is part of the German COVID-19 Research Network of University Medicine ("Netzwerk Universit\"atsmedizin”), funded by the German Federal Ministry of Education and Research (funding reference 01KX2021).}
}

\author{
	\IEEEauthorblockN{
		Felix Beierle\IEEEauthorrefmark{1},
		Uttam Dhakal\IEEEauthorrefmark{2},
		Caroline Cohrdes\IEEEauthorrefmark{3},
		Sophie Eicher\IEEEauthorrefmark{3},
		and
		R\"udiger Pryss\IEEEauthorrefmark{1}
	}
	\IEEEauthorblockA{
		\IEEEauthorrefmark{1}Institute of Clinical Epidemiology and Biometry,
		University of W\"urzburg,	W\"urzburg, Germany\\
		Email: felix.beierle@uni-wuerzburg.de, ruediger.pryss@uni-wuerzburg.de
	}
	\IEEEauthorblockA{\IEEEauthorrefmark{2}Service-centric Networking,
		Technische Universit\"at Berlin,
		Berlin, Germany\\
		Email: uttam.dhakal@campus.tu-berlin.de
	}
	\IEEEauthorblockA{\IEEEauthorrefmark{3}Department of Epidemiology and Health Monitoring, Robert Koch Institute,
		Berlin, Germany\\
		Email: cohrdesc@rki.de, eichers@rki.de
	}
}

\maketitle

\begin{abstract}

Several governments introduced or promoted
the use of contact-tracing apps during the ongoing COVID-19 pandemic.
In Germany, the related app is called \emph{Corona-Warn-App}, and by end of 2020,
it had 22.8 million downloads.
Contact tracing is a promising approach for containing the spread of
the novel coronavirus.
It is only effective if there is a large user base,
which brings new challenges like app users unfamiliar with using smartphones or apps.
As Corona-Warn-App is voluntary to use, reaching many users
and gaining a positive public perception is crucial for its
effectiveness.
Based on app reviews and tweets, we are analyzing the
public perception of Corona-Warn-App.
We collected and analyzed all 78,963 app reviews for the Android and iOS versions
from release (June 2020) to beginning of February 2021, as
well as all original tweets until February 2021 containing
\#CoronaWarnApp (43,082).
For the reviews, the most common words and n-grams point
towards technical issues, but it remains unclear,
to what extent this is due to the app itself,
the used \emph{Exposure Notification Framework},
system settings on the user's phone,
or the user's misinterpretations of app content.
For Twitter data, overall, based on tweet content, frequent hashtags,
and interactions with tweets,
we conclude that the German Twitter-sphere
widely reports adopting the app and promotes its use.

\end{abstract}


\section{Introduction}

The worldwide ubiquitous use of smartphones allows
for technical support in tackling the spread of the
novel coronavirus SARS-CoV-2 and the related ongoing COVID-19 pandemic
\cite{FerrettiQuantifyingSARSCoV2Transmission2020}.
Different countries with different cultures and different traditions regarding privacy and data protection implemented different solutions for contact-tracing apps.
Such apps mostly make sense when a significant part of the population uses them.
In order to achieve public acceptability of such apps,
there is the question of public perception and sentiment, i.e.,
what do the users of such an app and the general population think about it?
In this paper, we investigate the public perception of the contact-tracing app
used in Germany, Corona-Warn-App.
As an extension to existing survey-based studies
\cite{AltmannAcceptabilityAppBasedContact2020,HorstmannWhoDoesDoes2021},
we analyze app review data from the Apple App Store and from Google Play,
as well as public tweets with the corresponding hashtag \#CoronaWarnApp.
Our analysis can help to achieve a better understanding of the users' sentiment toward the app.
Our main contribution in this paper is the collection and analysis
of more than 75,000 app reviews and 40,000 tweets about the German COVID-19 contact-tracing app Corona-Warn-App.

The rest of the paper is structured as follows.
Section \ref{sec:rel-work} introduces the background
and details related work.
Section \ref{sec:method} describes the methodology and
how we collected and analyzed app reviews and tweets.
In Section \ref{sec:results}, we give the results,
which we discuss in Section \ref{sec:discussion},
before concluding in Section~\ref{sec:conclusion}.

\section{Background and Related Work}
\label{sec:rel-work}

\begin{figure}
	\center
	\includegraphics[width=0.725\columnwidth]{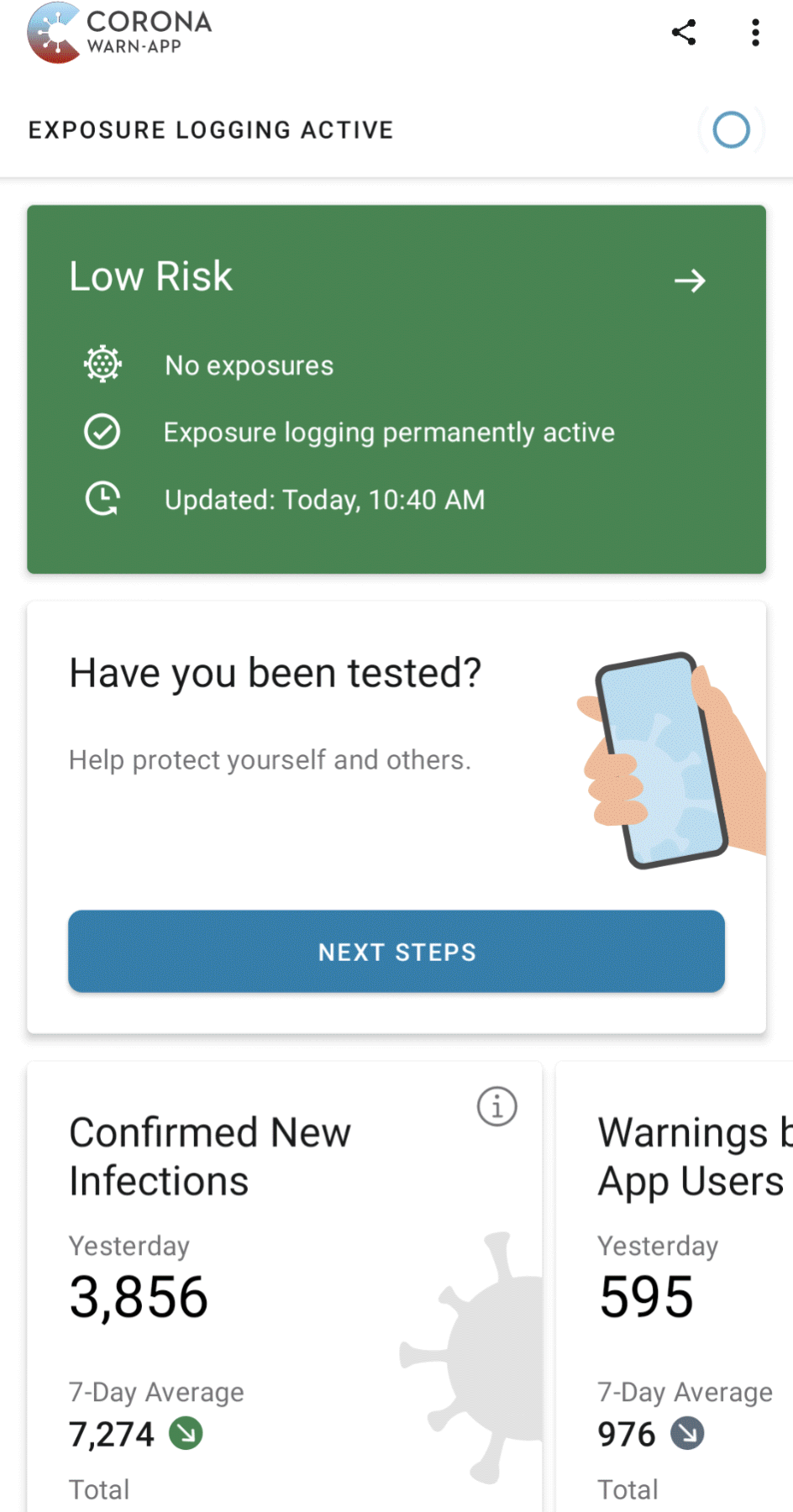}
	\label{fig:screenshot}
	\caption{Main screen of Corona-Warn-App.}
\end{figure}

Corona-Warn-App is an open source COVID-19 contact tracing app.
Figure \ref{fig:screenshot} shows the main screen of the app.
It was developed by Deutsche Telekom (one of the largest telecommunications companies is Europe)
and SAP (one of the largest software companies in the world) and released
by RKI (Robert Koch Institute), the German federal agency for public health responsible for disease control and prevention.
The app was released on June 16, 2020, its use is voluntary, and by the end of November 2020, 22.8 million
users downloaded the app, with Germany having a population of 83 million\footnote{\url{https://www.rki.de/DE/Content/InfAZ/N/Neuartiges_Coronavirus/WarnApp/Archiv_Kennzahlen/Kennzahlen_20112020.pdf}}.
The core functionality of Corona-Warn-App utilizes Bluetooth via the \emph{Exposure Notification Framework} from Apple/Google to trace which other devices have been in proximity.
The IDs that are exchanged are generated continually and are valid for 10-20 minutes.
They are derived from device-specific keys that are locally created every 24 hours.
The IDs that are marked as positive for a coronavirus test, are regularly downloaded
from a server and are locally compared to the list of devices met before\footnote{\url{https://www.coronawarn.app/en/}}.
When tested, the user can use the app to scan a QR-code at the doctor's office or testing station
and can get the result of the test via the app.
At the end of 2020, Corona-Warn-App introduced a new contact journal feature,
allowing users to manually log who they met.

Our previous research has shown that in general, users are quite willing to
share data with researchers when using apps voluntarily \cite{BeierleWhatdataare2020}.
Additionally, apps focusing on entertainment or social networking,
like Facebook or TikTok, have lots of users in Germany, without
concerns about privacy being a deal-breaking point for the general population.
The case of a governmental contact-tracing app is different in some key respects:
\begin{adjustwidth}{0.5cm}{}
\nointerlineskip\leavevmode
\paragraph*{\textbf{Target audience and user base}}	
The target audience or scope of the app is much higher compared to traditional apps
because of the ubiquity of the novel coronavirus and its effect on daily life.
Contact-tracing apps are not aimed at any specific age group or market segment, but are
aimed at anyone using a smartphone.
From the wider scope follows a more diverse user base than any other app.
This includes users that might be unfamiliar with app usage, smartphone usage, or technology
in general, and poses challenges to the design of the app.
\end{adjustwidth}

Because of the big scope of the app, Corona-Warn-App probably
was under special scrutiny from its inception.
Users might be more prone to be critical of the government,
combined with the large scope and user base, this might lead to
a large, critical user base with parts of it not having
the background to understand how the app works.
How the tracing technically works, what changing identifiers mean,
how data is exchanged, what data is actually exchanged, etc., is information
that is hard to convey to non-tech-savvy users.
All this can create potential misunderstandings
about technical limitations in terms of functionality
of the app.

Altmann et al.\
published a paper about a survey asking participants about contact-tracing apps
\cite{AltmannAcceptabilityAppBasedContact2020} (published end of August 2020).
They found very high willingness to install and use COVID-19 contact-tracing apps.
Horstmann et al.\ conducted a survey about the usage of Corona-Warn-App
\cite{HorstmannWhoDoesDoes2021} (published mid-December 2020).
A survey among 1,972 German adult reveals
that non-users had low trust in other people and often cited privacy concerns or doubts
regarding the app's effectiveness as reasons.
Now that a about a year of the pandemic has passed and such apps
are deployed and in use for some time, publicly available data can help
understand how users perceive them, rate them, and what aspects they
discuss about.

\section{Methodology}
\label{sec:method}

We used appbot\footnote{\url{https://appbot.co}} for exporting
publicly available app review data from the Apple App Store and Google Play.
We exported 78,963 reviews for the time until 4 February 2021.
Appbot detects the language; 75,891 of the reviews are in German (96\%).
Out of those, 77\% are for the Android app and 23\% for the iPhone app.
In contrast to only a \emph{rating}, a \emph{review} contains manual text input from the reviewing user.
For the same time window, appbot reports 166,439 ratings, 109,539 on Google Play (66\%), and 56,900 on Apple App Store (34\%).
This means that 47\% of users that rated the app also wrote a review.
We analyzed the ratings of the reviews, the reviews' distribution and ratings over time
as well as the review content.
We pre-processed the review texts, this includes stop word removal
and stemming, before we analyzed the review content in terms
of frequency of used words and n-grams.

Additionally, we collected all original tweets (i.e., no re-tweets) containing
\emph{\#CoronaWarnApp} until February 4, 2021 via the Twitter API v2.
Twitter has 5.45 million users in Germany as of end of 2020\footnote{\url{https://www.statista.com/statistics/242606/number-of-active-twitter-users-in-selected-countries/}}.
We collected 43,082 tweets
and analyzed the distribution and context of the tweets.
As with the reviews, we pre-processed the tweets before
looking into frequent words, n-grams, and hashtags.

\section{Results}
\label{sec:results}

First of all, we note that if the reported number of 22.8 million app downloads is correct,
only about 0.7\% of app downloaders rated the app, and about half of those left a review.
Still, the number of users reflected in our analysis is much higher than it is feasible in, e.g., a typical survey.
Furthermore, we assume that app reviews and tweets often
have a signaling effect because they can be posted by almost anyone
and are publicly visible.
Users that do not see their opinion expressed in the reviews or on Twitter can
just write their own review or tweet.

\subsection{App Reviews}
\label{sec:res-app-reviews}

The distribution between Android and iOS in the reviews (77\% Android) roughly matches that of the distribution of Android and iOS market shares in Germany (64\% Android)\footnote{\url{https://gs.statcounter.com/os-market-share/mobile/germany}}.
Looking at all ratings, including those without a review,
for Android, 36\% of the ratings were 5-star ratings, and 35\% were 1-star ratings.
The average was 3.0 stars.
For iOS, 75\% of the ratings were 5-star ratings, the average was 4.3 stars.
Just looking at the ratings that contain a manually written review, we get the results
shown in  Figures \ref{fig:android-review-ratings} and \ref{fig:ios-review-ratings}.
The averages for Android and iOS both were 2.9 stars.
Overall, the average rating of those with a review was lower, especially for iOS.
For the ratings with reviews, 1-star ratings made up the largest proportion
of the reviews, for both Android and iOS.
We consider this evidence that negative perceptions drive users
more to write a review than positive perceptions.
Figures \ref{fig:android-review-ratings} and \ref{fig:ios-review-ratings} show
that almost 70\% of the reviewers either gave a 1-star rating or a
5-star rating.
At the time of writing, other often downloaded apps like \emph{Jitsi Meet} or \emph{Zoom}
show a similar pattern of mostly 1-star or 5-star reviews.

\edef\mylst1{"20,831 (36\%)", "6,520 (11\%)", "6,912 (12\%)", "5,379 (9\%)", "18,839 (32\%)"}
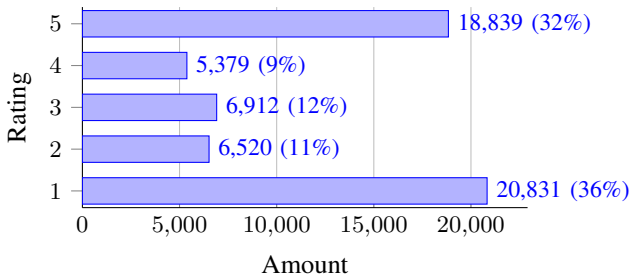
\begin{figure}[h]
	\begin{tikzpicture}
		\begin{axis}[
			width=7.5cm, height=4.25cm,
			every tick label/.append style={font=\small},
			xlabel={Amount},
			ylabel={Rating},
			xbar,
			xmin=0,
			ytick=data,
			axis x line*=bottom,
			axis y line*=left,
			scaled x ticks=false,
			xtick pos=left,
			ytick pos=left,
			xmajorgrids=true,
			every node near coord/.style={font=\small},
			nodes near coords=\pgfmathsetmacro{\mystring}{{\mylst1}[\coordindex]}\mystring,
			nodes near coords align={horizontal}]
			\addplot+[xbar] coordinates 
			{(20831,1) (6520,2) (6912,3) (5379,4) (18839,5)};
		\end{axis}
	\end{tikzpicture}
	\caption{Distribution of ratings from Android reviews.}
	\label{fig:android-review-ratings}
\end{figure}
\edef\mylst2{"6,354 (36\%)", "1,867 (11\%)", "2,134 (12\%)", "1,652 (9\%)", "5,403 (31\%)"}
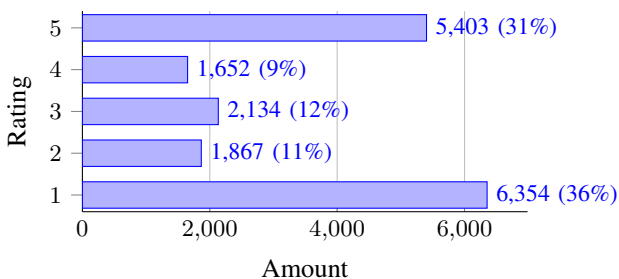
\begin{figure}[h]
	\begin{tikzpicture}
		\begin{axis}[
			width=7.5cm, height=4.25cm,
			every tick label/.append style={font=\small},
			xbar,
			xlabel={Amount},
			ylabel={Rating},
			xmin=0,
			ytick=data,
			axis x line*=bottom,
			axis y line*=left,
			scaled x ticks=false,
			xtick pos=left,
			ytick pos=left,
			xmajorgrids=true,
			every node near coord/.style={font=\small},
			nodes near coords=\pgfmathsetmacro{\mystring}{{\mylst2}[\coordindex]}\mystring,
			nodes near coords align={horizontal}]
			\addplot+[xbar] coordinates 
			{(6354,1) (1867,2) (2134,3) (1652,4) (5403,5)};
		\end{axis}
	\end{tikzpicture}
	\caption{Distribution of ratings from iOS reviews.}
	\label{fig:ios-review-ratings}
\end{figure}

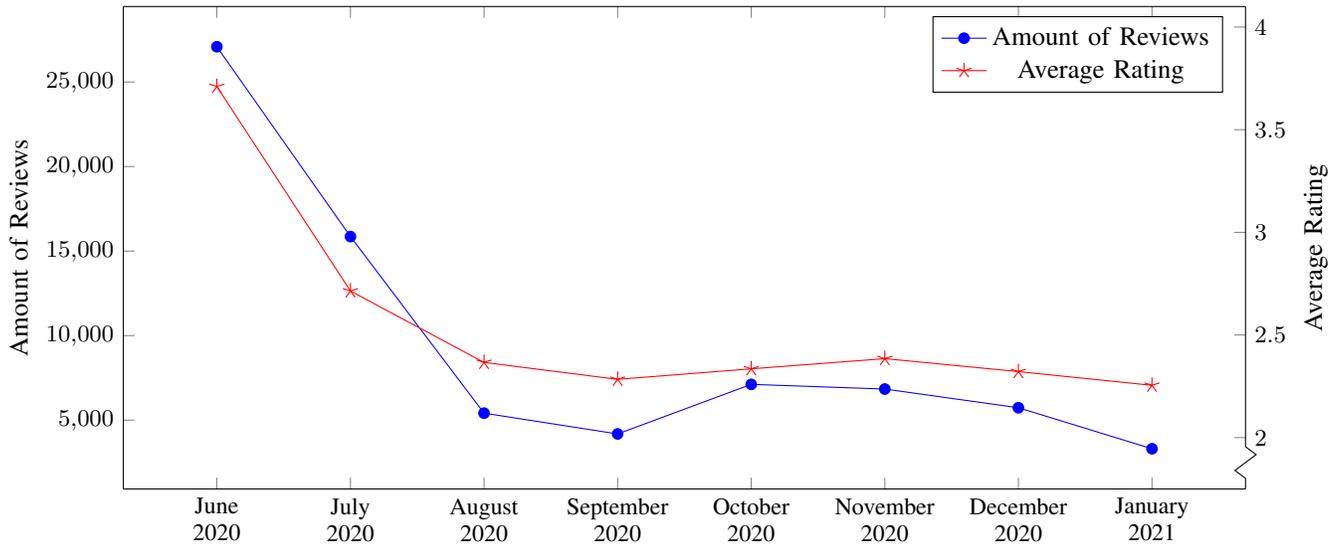
\begin{figure*}
\begin{tikzpicture}
	\begin{axis}[
		width=16.5cm, height=8.0cm,
		every tick label/.append style={font=\small},
		axis y line*=left,
		xtick=data,
		xticklabel style={align=center},
		xticklabels = {June\\2020, July\\2020, August\\2020, September\\2020, October\\2020, November\\2020, December\\2020, January\\2021},
		scaled y ticks=false,
		ylabel=Amount of Reviews]
		\addplot[blue, mark=*, mark options={scale=1, fill=blue}] coordinates 
		{(1,27085) (2, 15858) (3, 5414) (4, 4184) (5, 7116) (6, 6839) (7, 5733) (8, 3308)}; \label{plot_one}
	\end{axis}
	
	\begin{axis}[
		ymin=1.75,ymax=4.1,
		width=16.5cm, height=8.0cm,
		every tick label/.append style={font=\small},
		axis y line*=right,
		axis x line=none,
		axis y discontinuity=crunch,
		ylabel=Average Rating]
		\addlegendimage{/pgfplots/refstyle=plot_one}\addlegendentry{Amount of Reviews}
		\addlegendimage{/pgfplots/refstyle=plot_two}\addlegendentry{Average Rating}
		\addplot[red, mark=star, mark options={scale=1.5, fill=red}] coordinates 
		{(1, 3.710319) (2, 2.714340) (3, 2.36642) (4, 2.283939) (5, 2.335160) (6, 2.384705) (7, 2.321647) (8, 2.254837)}; \label{plot_two}
	\end{axis}
\end{tikzpicture}
\caption{Reviews over time and their average rating (Android and iOS combined).}
\label{fig:reviews-over-time}
\end{figure*}

Ratings of the reviews over time are shown in Figure \ref{fig:reviews-over-time}.
The blue line with circles shows the amount of downloads
and the red line with stars shows the average number of ratings.
We observe that 36\% of all reviews (as of end of January 2020) were from June 2020.
The amount of reviews was in steady decline,
except for a slight increase in October 2020.
The average rating in the reviews started out at 3.6 in June 2020,
declined rapidly after that, and stayed at around 2.4 since August 2020.

For Google Play, appbot exports the public reply rates of the app developer.
The provider of the app can directly answer to reviews.
The was done to a large extent for Corona-Warn-App.
The reply rate was 50\%-60\% for all reviews with 1 to 4 stars,
and 31\% for the reviews with 5 stars.
There were 26,138 manual replies to reviews,
indicating a large effort taken.

In order to get a better understanding of what users are writing about in their reviews,
we looked into the most frequent words and n-grams.
Taking into account the 75,891 German reviews,
for pre-processing, we removed stop words.
Then, we used the CISTEM stemmer for getting the stem of each word \cite{WeissweilerDevelopingStemmerGerman2018}.

Table \ref{tbl:freq-words-reviews} shows the English translation of the most frequent word stems
in the German reviews.
Some of the words can be used in both positive as well as negative contexts,
for example, a review could contain "The app is not \emph{good}" or "The app is \emph{good}."
The word \emph{good} alone does not suffice to interpret the sentiment of the review.
There are some words, however, that we regard as relevant for understanding
the public perception of the app.
We assume that 
\emph{(to) function}, \emph{update}, \emph{test}, and \emph{error}
all refer to the functionality of the app.
While \emph{update} and \emph{error} likely refer to the technical side,
\emph{test} likely refers to coronavirus tests, for which
the result can be received via the app.

\begin{table}[h]
	\small
	\centering
	\caption{Most frequent words in app reviews (Android and iOS).}
	\label{tbl:freq-words-reviews}
	\begin{tabular}{@{}lr@{}}
		\toprule
		\textbf{Word} & \textbf{Occurrences} \\ 
		day & 20,137 (27\%)	\\ \hdashline[0.5pt/2.5pt]
		to function & 15,552 (20\%)	\\ \hdashline[0.5pt/2.5pt]
		more & 13,116 (17\%)	\\ \hdashline[0.5pt/2.5pt]
		since & 12,873 (17\%)	\\ \hdashline[0.5pt/2.5pt]
		good & 12,237 (16\%)	\\ \hdashline[0.5pt/2.5pt]
		always & 9,241 (12\%)	\\ \hdashline[0.5pt/2.5pt]
		update & 8,707 (11\%)	\\ \hdashline[0.5pt/2.5pt]
		test & 8,705 (11\%)	\\ \hdashline[0.5pt/2.5pt]
		error & 8,462 (11\%)	\\ \bottomrule
	\end{tabular}
\end{table}

Table \ref{tbl:bigrams-reviews} shows the most frequent bigrams,
and Table \ref{tbl:trigrams-reviews} shows the most frequent trigrams.
We observe that \emph{day} was the most frequent word
and that there were two frequent bigrams containing "\emph{day}":
\emph{since, day} and \emph{day, active}.
When Corona-Warn-App is started for the first time, it indicates that
because of the incubation period, the app should be used for 14 days to track encounters for long enough before meaningful results can be displayed.
For this time period, the number of days the app is already in use is shown on the main screen.
\emph{Day, active} probably refers to this.
\emph{Since, day} might refer to the same aspect, or to some other time period in which the user is waiting for something, for example, change in app functionality, or waiting for a test result.
All other bigrams and all listed trigrams refer to the functionality of the app.
There seems to be a focus on technical aspects: Google, API, QR Code, communication.

\begin{table}[]
	\small
	\centering
	\caption{Most frequent bigrams in app reviews (Android and iOS).}
	\label{tbl:bigrams-reviews}
	\begin{tabular}{@{}lr@{}}
		\toprule
		\textbf{Bigram} & \textbf{Occurrences} \\ 
since, day & 3,191 (4\%)	\\ \hdashline[0.5pt/2.5pt]
positive, test & 1,675 (2\%)	\\ \hdashline[0.5pt/2.5pt]
Google, API & 1,572 (2\%)	\\ \hdashline[0.5pt/2.5pt]
new, to install & 1,451 (2\%)	\\ \hdashline[0.5pt/2.5pt]
QR, code & 1,333 (2\%)	\\ \hdashline[0.5pt/2.5pt]
communication, Google & 1,276 (2\%) \\ \hdashline[0.5pt/2.5pt]
error, communication & 1,240 (2\%) \\ \hdashline[0.5pt/2.5pt]
day, active & 1,216 (2\%) \\ \hdashline[0.5pt/2.5pt]
last, update & 1,185 (2\%) \\ \bottomrule
	\end{tabular}
\end{table}

\begin{table}[]
	\small
	\centering
	\caption{Most frequent trigrams in app reviews (Android and iOS).}
	\label{tbl:trigrams-reviews}
	\begin{tabular}{@{}lr@{}}
		\toprule
		\textbf{Trigram} & \textbf{Occurrences} \\ 
error, communication, Google & 1,125 (1\%)	\\ \hdashline[0.5pt/2.5pt]
communication, Google, API & 1,003 (1\%)	\\ \hdashline[0.5pt/2.5pt]
since, last, update & 629 (1\%)	\\ \hdashline[0.5pt/2.5pt]
cause, gone, wrong & 423 (1\%) \\ \bottomrule
	\end{tabular}
\end{table}

\subsection{Twitter}
\label{sec:twitter}

The first tweet with \emph{\#CoronaWarnApp} appeared on April 2, 2020.
We collected all original tweets containing this hashtag until February 4, 2021.
There were 43,082 tweets overall.
In Figure \ref{fig:tweets-over-time}, we show the number of tweets
per day.
Between June 15, 2020 and June 19, 2020, there were 18,901 tweets (42\% of all tweets),
with a maximum of 12,058 tweets (28\% of all tweets) on June 16, 2020.
Even more than for the reviews, the peak in terms of number of tweets
immediately followed after the initial release.

After that, the plot in Figure \ref{fig:tweets-over-time} shows
peaks on July 24, August 4, and September 23.
For July 24, the most common words and hashtags did not show any
specific topic besides those covered at other times.
For August 4, one single, unverified user
tweeted 222 original tweets, containing spam.
On September 23, among the frequent words and hashtags was "Illner,"
the name of a talk show on German TV.
Likely, the COVID-19 pandemic and Corona-Warn-App
were topics in the show and sparked some statements or discussions on Twitter.
At the end of October 2020, Germany saw new record numbers
of new infections several times.
In November, new restrictions to public life were introduced.
We assume that the increase in tweets at the end of October can be related to that.

	\pgfplotsset{
	every non boxed x axis/.style={} 
	}

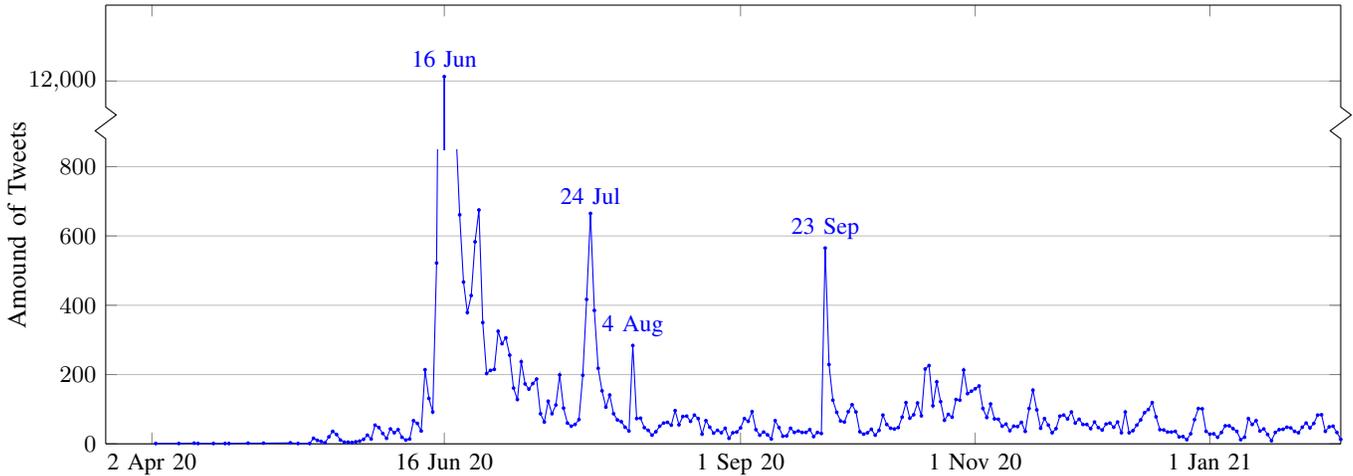
\begin{figure*}
	\setlength\abovecaptionskip{-16.0\baselineskip}
	\begin{tikzpicture}
	\begin{groupplot}[
		group style={
			group size=1 by 2,
			horizontal sep=0pt,
			vertical sep=0pt
		},
		width=18.0cm,
		ymajorgrids,
		scaled y ticks=false,
		point meta=explicit symbolic,
		every node near coord/.style={font=\small},
		every tick label/.append style={font=\small},
		groupplot ylabel={Amound of Tweets}
		]
		
		\nextgroupplot[ymin=11100,ymax=13000,
		ytick={12000},
		axis x line=top,
		axis y discontinuity=crunch,
		date coordinates in=x,
		xmin={2020-03-20},
		xmax={2021-02-04},
		xtick={2020-04-01, 2020-06-16, 2020-09-01, 2020-11-01, 2021-01-01},
		xticklabels={},
		xticklabel style={align=center},
		scaled y ticks=false,
		point meta=explicit symbolic,
		height=3.5cm]
		\addplot[nodes near coords, blue, mark=*, mark options={scale=0.25, fill=blue}] table [col sep=comma,trim cells=true,x=date,y=amount,meta=label] {tweets-per-day2-group2-v3.txt};     
		
		\nextgroupplot[ymin=0,ymax=850,
		ytick={0,200,400,600,800},
		axis x line=bottom,
		date coordinates in=x,
		xmin={2020-03-20},
		xmax={2021-02-04},
		xtick={2020-04-01, 2020-06-16, 2020-09-01, 2020-11-01, 2021-01-01},
		xticklabels={2 Apr 20, 16 Jun 20, 1 Sep 20, 1 Nov 20, 1 Jan 21},
		xticklabel style={align=center},
		scaled y ticks=false,
		point meta=explicit symbolic,
		height=5.5cm
		]
		\addplot[nodes near coords, blue, mark=*, mark options={scale=0.25, fill=blue}] table [col sep=comma,trim cells=true,x=date,y=amount,meta=label] {tweets-per-day2-group1-v3.txt};               
	\end{groupplot}
	\end{tikzpicture}
	\caption{Original tweets per day with \#CoronaWarnApp.}
	\label{fig:tweets-over-time}
\end{figure*}

Looking at the top 10 tweet groups
\emph{most retweets}, \emph{most likes}, and \emph{most replies} of German tweets,
we observe that 9 out of 10 tweets were from verified accounts.
The average follower count in each group was >400,000.
There was some overlap of the users in each group.
Overall, the users were mostly
publicly funded television programs,
journalists, politicians, and scientists.

For the 36,304 German tweets (84\% of all tweets),
we did the same pre-processing of stop word removal and stemming with CISTEM as for the reviews.
Some of the most frequent words from the reviews
(see Table \ref{tbl:freq-words-reviews})
were also most frequent words in the tweets:
\emph{to function},
\emph{more},
\emph{good},
\emph{day},
\emph{always},
\emph{since},
\emph{test}.
They occurred in 3\% to 10\% of the tweets.
\emph{To install} was the most frequent word in tweets
and was present in 10\% of the German tweets.
We note that neither \emph{update}, nor \emph{error} was among
the most frequent words in the tweets.
\emph{Data protection}, not among the frequent words
in the reviews, appeared in 1,708 (5\%) of the tweets\footnote{Note that \emph{data protection} is one word in German: \emph{Datenschutz}.}.
In their work, Xue et al.\ analyzed
4 million tweets related to COVID-19 and found \emph{virus}
to be the most frequent word with a frequency of 1.18\% \cite{XueTwitterDiscussionsEmotions2020}.
We assume that the narrower topic of Corona-Warn-App
leads to a smaller variety of tweet contents,
i.e., we would expect tweets with \#CoronaWarnApp to
vary less in content compared to tweets about the coronavirus in general.
This, in turn, means that tweets with \#CoronaWarnApp
are likely more similar to each other compared to tweets from
a broader topic, leading to the much higher frequencies
of words we observed.

Next, we looked into the hashtags used in tweets containing \#CoronaWarnApp.
Leaving out those directly related to corona (like \#Corona, \#COVID19, etc.),
and disregarding the case of the letters,
the most common hashtags are given in Table \ref{tbl:freq-hashtags-tweets}.
The content here is much less focused on technical aspects.
The names of politicians,
terminology commonly associated with political debate,
like \emph{mandatory mask-wearing}, \emph{data protection}, \emph{covidiot}, or \emph{basic rights} are evidence for
rather political than technical tweets.
Additionally, \#IchAppMit, which uses \emph{app} as a verb and signals
being part of the group that uses Corona-Warn-App,
indicates promoting using the app.

\begin{table}[]
	\small
	\centering
	\caption{Most frequent hashtags in original tweets with \#CoronaWarnApp, disregarding hashtags directly related to the virus.}
	\label{tbl:freq-hashtags-tweets}
	\begin{tabular}{@{}llr@{}}
		\toprule
		\textbf{Hashtag} & \textbf{Translation/Explanation} & \textbf{Occurrences} \\ 
\#Datenschutz & data protection & 831 (2.3\%)	\\ \hdashline[0.5pt/2.5pt]
\#IchAppMit & roughly: I am also using the app & 777 (2.1\%)	\\ \hdashline[0.5pt/2.5pt]
\#Maskenpflicht & mandatory mask-wearing & 416 (1.1\%)	\\ \hdashline[0.5pt/2.5pt]
\#RKI & Robert Koch Inst., see Section \ref{sec:rel-work} & 381 (1.0\%)	\\ \hdashline[0.5pt/2.5pt]
\#Spahn & German federal minister of health & 376 (1.0\%) \\ \hdashline[0.5pt/2.5pt]
\#Merkel & German chancellor & 320 (0.9\%)	\\ \hdashline[0.5pt/2.5pt]
\#App & & 317 (0.9\%) \\ \hdashline[0.5pt/2.5pt]
\#Covidioten & \emph{covidiots} & 307 (0.9\%)	\\ \hdashline[0.5pt/2.5pt]
\#Grundrechte & basic rights & 255 (0.7\%)	\\ \bottomrule
	\end{tabular}
\end{table}

\section{Discussion}
\label{sec:discussion}

First, we note the low rate of reviews compared to the number of downloads.
As app download numbers are not publicly available
in the Apple App Store, and Google Play only reports broad ranges of download numbers,
we look at a different area for comparison and interpretability.
For measuring success on YouTube,
successful videos gain a comments-to-view ratio
of 0.5\%\footnote{\url{https://tubularlabs.com/blog/3-metrics-youtube-success/}},
which is similar to the review-to-download ratio (0.7\%) for Corona-Warn-App.
Maybe using this metric as a metric for success does not translate
from YouTube to app reviews,
but the comparison helps explaining the seemingly low interaction rate.

Given that 36\% of all reviews were from June 2020, the month of release of the app,
it is likely that the users posting those very early reviews
with relatively high ratings (see Figure \ref{fig:reviews-over-time})
did not in fact use the app for very long, but
used a positive app review as a means of supporting or promoting the app.

Looking at the average rating for the Android and the iOS app, we note a big difference (3.0 vs.\ 4.3).
On the basis of those numbers only, it is difficult to draw conclusions about the reasons.
One factor could be the variety of Android devices and the associated challenges for development.
In an academic project,
we released the Android app TYDR \cite{BeierleTYDRTrackYour2018a,BeierleContextDataCategories2018} and with around 4,000 installations, we already identified more than 600 different devices.

From Google Play, for Corona-Warn-App, we could see the high response rates of the app developer of up to 60\%.
We interpret this as strong engagement with the users,
to signal taking the reviews seriously and
replying to questions, problems, and concerns.

Looking at the frequent words and n-grams in the reviews, we noted
a focus on technical terms.
We interpret this as users reporting about (issues in) their usage of the app.
There could be several factors at play
that have lead to the large amount of negative reviews and the technical focus of the reviews.
Users without proficiency in technology
were likely not be able to distinguish between bugs in the app,
(temporary) problems with the underlying Google/Apple \emph{Exposure Notification Framework}, or system-wide settings, e.g., regarding Bluetooth or notifications.
Thus, for example, it is hard to tell if a user that reported a problem
ran into a well-known issue in the Android ecosystem, like the
reliability of background processing,
if he/she ran into a bug of Corona-Warn-App, (accidentally) turned his/her Bluetooth off,
or did not understand how to interpret what is shown on the screen.

Another factor regarding the public perception of the app
might be the world's reaction to the pandemic,
with lockdowns and social distancing that likely no-one alive
has ever seen to that extent before.
Uncertainty and anxiety are likely widespread and might fuel
dissatisfaction with Corona-Warn-App independent of its functionality and design.
Some "legacy" real-world systems are integrated into Corona-Warn-App,
e.g., getting the result of a test for the novel coronavirus.
Waiting for the results might be frustrating
and writing a review might be an easy outlet for that frustration.

Given the broad user base, and our hypotheses about different
root-causes of the public perception of Corona-Warn-App,
there might be multiple routes for future work.
There might be technical problems within the app, and/or with
the \emph{Exposure Notification Framework}.
Additionally, public education about technology might increase
proper usage and help avoid confusion when using apps like Corona-Warn-App.
Another field to research is the UX/UI of the app
with the broad target audience and user base in mind.
Lastly, the improvement of the integration of "legacy" real-world systems
likely progresses slowly because laws and regulation might have to be changed.

Looking into the data we collected from Twitter, we firstly note that
the user base being active on Twitter in Germany is much smaller than
the number of people who downloaded the app.
The volume of tweets per day already gives evidence about their content.
28\% of all original tweets were published on June 16, 2020, see Section \ref{sec:twitter}.
We regard this as evidence that a large number of tweets
might be news-related.
The most frequent words and hashtags indicate
that Twitter users are interested in political topics.
Overall, given the data we analyzed, in the Twitter-sphere,
the overall sentiment is a positive one towards using the app.

There has been work analyzing the spread of
misinformation \cite{GallottiAssessingRisksInfodemics2020}
and conspiracy theories
\cite{AhmedCOVID195GConspiracy2020,AhmedCOVID19FilmYour2020}
during the COVID-19 pandemic.
Based on the most frequent words, bigrams, and hahstags in the reviews
and tweets, we find no evidence of wide-spread conspiracy theories
related to Corona-Warn-App.
Note that one limiting factor in drawing this conclusion is
that Google, Apple, and Twitter might have already deleted
such posts at the time we collected our data (beginning of 2021).

\section{Conclusions and Future Work}
\label{sec:conclusion}

We collected and analyzed 78,963 app reviews about Corona-Warn-App
and 43,083 original tweets containing \emph{\#CoronaWarnApp}.
In the reviews, the users seem mostly interested in the apps'
functionality, while Twitter was mostly used
for news and promoting awareness of the app.

The reviews contain technical terms, often reporting issues.
It remains open to what extent these reports are
based on bugs in the app itself,
on errors in the underlying OS or used libraries, specifically the \emph{Exposure Notification Framework},
or on misunderstandings by the user.
The especially diverse user base, potentially containing users
that are not familiar with using smartphones and apps,
poses special challenges regarding
the design of such an app.

Overall, both the timeline of tweets
as well as the most frequent words and hashtags
indicate that Twitter is being used for spreading news
about the app and promoting its use.
One limitation of our work is that both data sources are likely
biased and thus cannot be considered representative for the German population.
The reviews are only from people who downloaded the app,
and the Twitter-sphere represents only a fraction of the German population.

Future work includes
investigating how to better convey using apps to
a non-tech-savvy audience.
Additionally, future work should focus on the users' motivations to use the app
and their acceptance of the app.
Such insights can help to address the users' needs and expectations.
This, in turn, could help to increase compliance with and efficacy of targeted prevention measures in the long term.

\bibliographystyle{IEEEtran}

\begin{thebibliography}{10}
\providecommand{\url}[1]{#1}
\csname url@samestyle\endcsname
\providecommand{\newblock}{\relax}
\providecommand{\bibinfo}[2]{#2}
\providecommand{\BIBentrySTDinterwordspacing}{\spaceskip=0pt\relax}
\providecommand{\BIBentryALTinterwordstretchfactor}{4}
\providecommand{\BIBentryALTinterwordspacing}{\spaceskip=\fontdimen2\font plus
\BIBentryALTinterwordstretchfactor\fontdimen3\font minus
  \fontdimen4\font\relax}
\providecommand{\BIBforeignlanguage}[2]{{%
\expandafter\ifx\csname l@#1\endcsname\relax
\typeout{** WARNING: IEEEtran.bst: No hyphenation pattern has been}%
\typeout{** loaded for the language `#1'. Using the pattern for}%
\typeout{** the default language instead.}%
\else
\language=\csname l@#1\endcsname
\fi
#2}}
\providecommand{\BIBdecl}{\relax}
\BIBdecl

\bibitem{FerrettiQuantifyingSARSCoV2Transmission2020}
L.~Ferretti, C.~Wymant, M.~Kendall, L.~Zhao, A.~Nurtay, L.~{Abeler-D{\"o}rner},
  M.~Parker, D.~Bonsall, and C.~Fraser, ``Quantifying {{SARS}}-{{CoV}}-2
  transmission suggests epidemic control with digital contact tracing,''
  \emph{Science}, vol. 368, no. 6491, May 2020.

\bibitem{AltmannAcceptabilityAppBasedContact2020}
S.~Altmann, L.~Milsom, H.~Zillessen, R.~Blasone, F.~Gerdon, R.~Bach,
  F.~Kreuter, D.~Nosenzo, S.~Toussaert, and J.~Abeler, ``Acceptability of
  {{App}}-{{Based Contact Tracing}} for {{COVID}}-19: {{Cross}}-{{Country
  Survey Study}},'' \emph{JMIR mHealth and uHealth}, vol.~8, no.~8, p. e19857,
  Aug. 2020.

\bibitem{HorstmannWhoDoesDoes2021}
K.~T. Horstmann, S.~Buecker, J.~Krasko, S.~Kritzler, and S.~Terwiel, ``Who does
  or does not use the `{{Corona}}-{{Warn}}-{{App}}' and why?'' \emph{European
  Journal of Public Health}, vol.~31, no.~1, pp. 49--51, Feb. 2021.

\bibitem{BeierleWhatdataare2020}
F.~Beierle, V.~T. Tran, M.~Allemand, P.~Neff, W.~Schlee, T.~Probst,
  J.~Zimmermann, and R.~Pryss, ``What data are smartphone users willing to
  share with researchers?'' \emph{Journal of Ambient Intelligence and Humanized
  Computing}, vol.~11, pp. 2277--2289, 2020.

\bibitem{WeissweilerDevelopingStemmerGerman2018}
L.~Weissweiler and A.~Fraser, ``Developing a {{Stemmer}} for {{German Based}}
  on a {{Comparative Analysis}} of {{Publicly Available Stemmers}},'' in
  \emph{Language {{Technologies}} for the {{Challenges}} of the {{Digital
  Age}}}, ser. Lecture {{Notes}} in {{Computer Science}}, G.~Rehm and
  T.~Declerck, Eds.\hskip 1em plus 0.5em minus 0.4em\relax {Springer
  International Publishing}, 2018, pp. 81--94.

\bibitem{XueTwitterDiscussionsEmotions2020}
J.~Xue, J.~Chen, R.~Hu, C.~Chen, C.~Zheng, Y.~Su, and T.~Zhu, ``Twitter
  {{Discussions}} and {{Emotions About}} the {{COVID}}-19 {{Pandemic}}:
  {{Machine Learning Approach}},'' \emph{Journal of Medical Internet Research},
  vol.~22, no.~11, p. e20550, 2020.

\bibitem{BeierleTYDRTrackYour2018a}
F.~Beierle, V.~T. Tran, M.~Allemand, P.~Neff, W.~Schlee, T.~Probst, R.~Pryss,
  and J.~Zimmermann, ``{{TYDR}} \textendash{} {{Track Your Daily Routine}}.
  {{Android App}} for {{Tracking Smartphone Sensor}} and {{Usage Data}},'' in
  \emph{2018 {{IEEE}}/{{ACM}} 5th {{International Conference}} on {{Mobile
  Software Engineering}} and {{Systems}} ({{MOBILESoft}})}.\hskip 1em plus
  0.5em minus 0.4em\relax {ACM}, 2018, pp. 72--75.

\bibitem{BeierleContextDataCategories2018}
------, ``Context {{Data Categories}} and {{Privacy Model}} for {{Mobile Data
  Collection Apps}},'' \emph{Procedia Computer Science}, vol. 134, pp. 18--25,
  2018.

\bibitem{GallottiAssessingRisksInfodemics2020}
R.~Gallotti, F.~Valle, N.~Castaldo, P.~Sacco, and M.~De~Domenico, ``Assessing
  the risks of `infodemics' in response to {{COVID}}-19 epidemics,''
  \emph{Nature Human Behaviour}, vol.~4, no.~12, pp. 1285--1293, Dec. 2020.

\bibitem{AhmedCOVID195GConspiracy2020}
W.~Ahmed, J.~{Vidal-Alaball}, J.~Downing, and F.~L. Segu{\'i}, ``{{COVID}}-19
  and the {{5G Conspiracy Theory}}: {{Social Network Analysis}} of {{Twitter
  Data}},'' \emph{Journal of Medical Internet Research}, vol.~22, no.~5, p.
  e19458, May 2020.

\bibitem{AhmedCOVID19FilmYour2020}
W.~Ahmed, F.~L. Segu{\'i}, J.~{Vidal-Alaball}, and M.~S. Katz, ``{{COVID}}-19
  and the ``{{Film Your Hospital}}'' {{Conspiracy Theory}}: {{Social Network
  Analysis}} of {{Twitter Data}},'' \emph{Journal of Medical Internet
  Research}, vol.~22, no.~10, p. e22374, 2020.

\end{thebibliography}

\end{document}